\documentclass[fleqn,12pt,twoside]{article}
\usepackage{espcrc1}

\usepackage{graphicx}

\newcommand{\AmS}{{\protect\the\textfont2
  A\kern-.1667em\lower.5ex\hbox{M}\kern-.125emS}}

\title{Wavelets in Momentum-Space Scattering Calculations}

\author{B. M. Kessler, G. L. Payne, and W. N. Polyzou
        \address{Department of Physics and Astronomy\\ 
        The University of Iowa \\ 
        Iowa City, IA, 52246}%
\thanks{This work  was supported in part by 
the Department of Energy, Nuclear Physics Division, under contract DE-FG02-86ER40286.}}
       
\begin{document}

\maketitle

\begin{abstract}

We demonstrate that wavelet bases have features that make them
advantageous for solving momentum-space scattering integral equations.
Using the example of two nucleons interacting with the Malfliet-Tjon V
interaction, we show it is possible to reduce the size of the matrix
representation of the Lippmann-Scwhinger equation by 96\% with a loss
of accuracy of only a few parts in a million.  We also demonstrate
that wavelet methods provide an accurate means for treating the
singularities that appear in the scattering integral equations.

\end{abstract}

\bigskip

We show that the use of Daubechies' \cite{daub1} wavelet bases for
solving momentum-space scattering integral equations \cite{bk1} leads
to sparse matrices which simplify the numerical solution.  The method
is tested on nucleon-nucleon scattering with a Malfliet-Tjon
V \cite{mj} potential.  The results of the test calculations indicate
that a significant reduction in computational size can be achieved for
realistic few-body scattering calculations.
				   
The motivation for this work is that accurate relativistic few-body
calculations are important for the few-body program at TJNAF.  These
calculations are naturally formulated in momentum-space because
of the need to include momentum-dependent Wigner and/or Melosh
rotations, and the nonlinear relation between mass and energy in
relativistic theories.  The computational disadvantage of 
momentum-space scattering integral equations is that the kernels are
represented by large dense matrices and the computation of the matrix
elements involves singular integrals.  Our work indicates that if the
solutions of the integral equations are represented as expansions in a
wavelet basis then the resulting kernel can be expressed as the sum of a
sparse matrix and a matrix with small norm.  Setting the small matrix to zero
leads to an accurate approximation.   The density of the non-zero 
matrix elements of the sparse matrix representation of the 
s-wave Malfliet-Tjon V K-matrix kernel is illustrated in the 
two-dimensional plot in fig 1.
\begin{figure}[hb]
\centering
\includegraphics[width=3.5in]{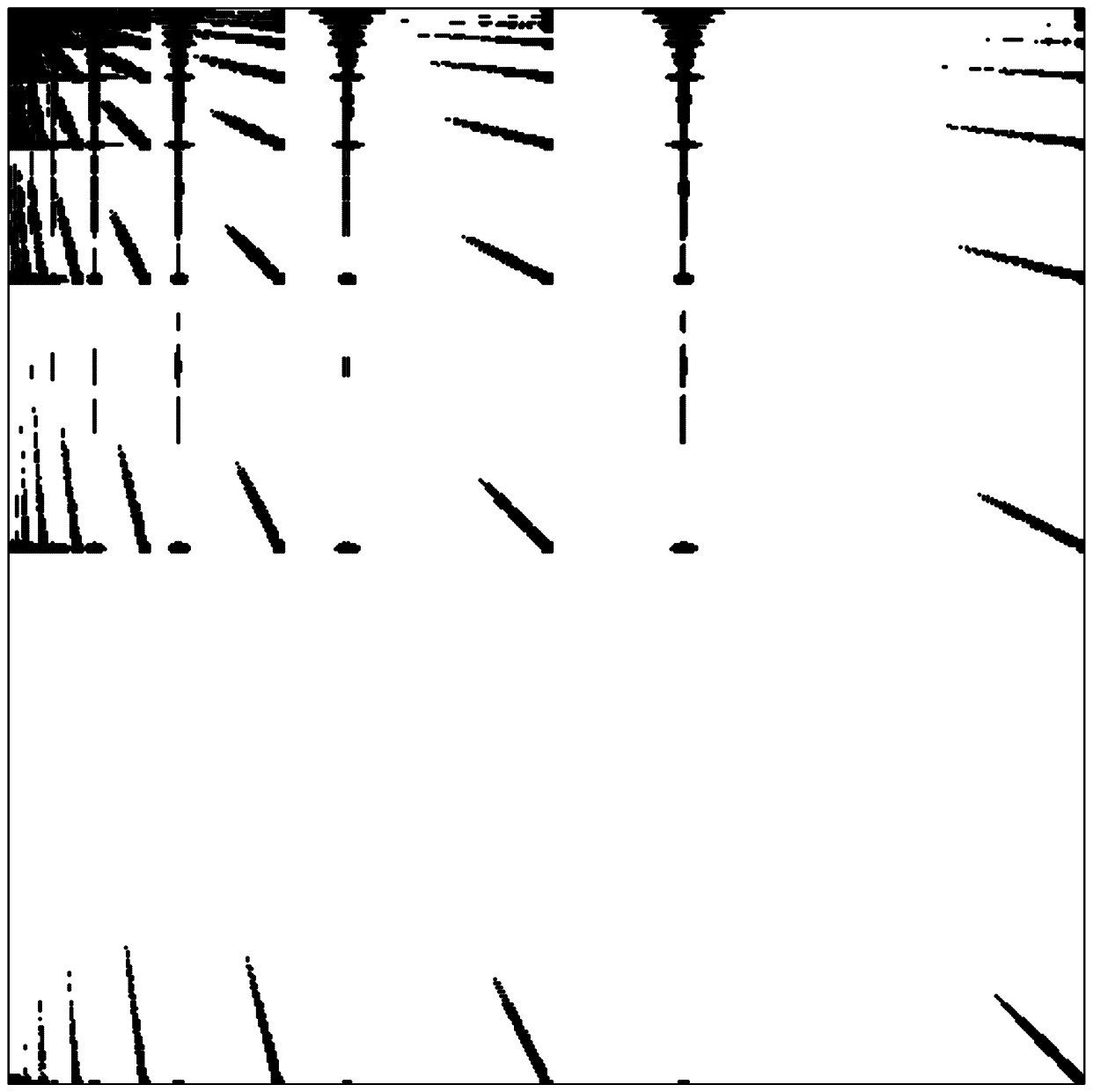}
\caption{Non-zero matrix elements of $s$-wave kernel of the $K$-matrix 
equation for the Malfliet Tjon V potential 
in the Daubechies wavelet basis of order three.}
\label{fig1}
\end{figure}

The Daubechies wavelets have many features of a spline basis; the
basis functions have compact support and finite linear combinations of
the basis functions can (pointwise) locally represent polynomials.  Unlike the
spline basis, the wavelet basis functions are orthonormal, and there
is an automatic method for determining the most important basis
elements for a given calculation.  The wavelet basis functions also
have natural quadrature rules that lead to efficient methods for
calculating kernel matrix elements. There is a robust wavelet-based
method to accurately and efficiently compute the singular integrals
that arise in momentum space-scattering calculations.

While wavelets are used extensively in signal processing, they have not
been used as much in numerical analysis.
Part of the reason for this is that the wavelet
basis is related by a simple orthogonal transformation to the solution
of a linear renormalization group equation.  The solution of this
equation, which is called the scaling function, 
has a fractal structure, having structure on all scales.
This fractal structure also appears in the wavelet basis functions and
limits the applicability of numerical methods which assume that
functions are smooth on a sufficiently small scale.  The structure of
the Daubechies wavelet and scaling function 
of order three are shown in fig 2.
\begin{figure}[hb]
\centering
\includegraphics[width=3.5in]{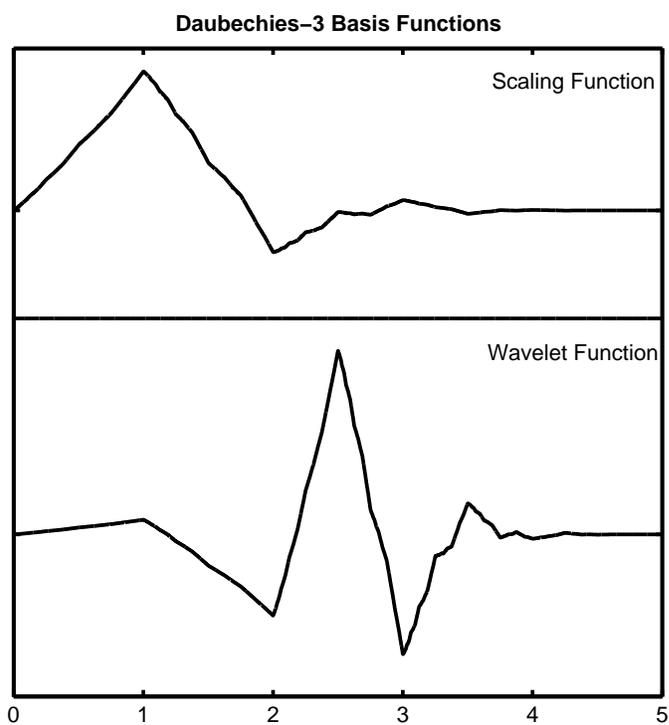}
\caption{Plot of the Daubechies order-three wavelet and scaling function.}
\label{fig2}
\end{figure}
Nevertheless, we show that the renormalization group equation provides
a robust tool for developing alternative methods of numerical
computation which overcome all of the difficulties that result from
the fractal structure \cite{notes}.

Our calculation of the $K$-matrix for a Malfliet-Tjon interaction has
the property that if the smallest 96\% of the matrix elements in the
kernel of the numerical representation of the Lippmann-Schwinger
equation are set to zero, the mean square error is only a few parts in
a million.  This allows for the replacement of the kernel matrix by a
sparse matrix.  We have also developed methods to limit the amount of
storage necessary to transform the original equation to this
sparse-matrix representation.  These preliminary results strongly
suggest that wavelet methods have many advantages for the
numerical treatment of momentum space scattering equations.

\end{document}